\documentclass{eptcs}

\def\event{PLACES 2013}

\usepackage{amssymb}
\usepackage{stmaryrd}
\usepackage{proof}

\newcommand{\stmt}{\mathit{stmt}}

\newcommand{\lpar}{\mathbin{\llfloor}}
\newcommand{\rpar}{\mathbin{\rrfloor}}
\newcommand{\Atomic}[1]{\mathsf{atomic}~#1}
\newcommand{\Await}[2]{\mathsf{await}~#1~\mathsf{do}~#2}
\newcommand{\Suspend}{\mathsf{suspend}}

\newcommand{\Skip}{\ensuremath{\mathsf{skip}}}
\newcommand{\Assign}[2]{\ensuremath{#1 := #2}}
\newcommand{\Seq}[2]{\ensuremath{#1;#2}}
\newcommand{\Ifthenelse}[3]
{\ensuremath{\mathsf{if~} #1 \mathsf{~then~} #2 \mathsf{~else~} #3}}
\newcommand{\While}[2]{\ensuremath{\mathsf{while~} #1 \mathsf{~do~} #2}}
\newcommand{\true}{\ensuremath{\mathsf{true}}}

\newcommand{\istrue}[2]{\ensuremath{#1 \models #2}}
\newcommand{\isfalse}[2]{\ensuremath{#1 \not\models #2}}
\newcommand{\update}[3]{\ensuremath{#1[#2\mapsto #3]}}

\newcommand{\res}{\mathit{res}}

\newcommand{\gres}{\mathit{res}_\mathrm{g}}
\newcommand{\xcfg}{\mathit{xcfg}}

\newcommand{\red}[3]{#1, #2 \to #3}
\newcommand{\mmred}[3]{#1, #2 \to^\mathrm{m} #3}
\newcommand{\gmmred}[3]{#1, #2 \to^\mathrm{m}_\mathrm{g} #3}

\newcommand{\ret}[1]{\mathit{ret}~#1}
\newcommand{\delay}[1]{\delta~#1}

\newcommand{\yield}[1]{\mathit{yield}~#1}

\newcommand{\bism}[2]{#1 \sim #2}
\newcommand{\gbism}[2]{#1 \mathrel{\sim_\mathrm{g}} #2}

\newcommand{\exec}[3]{ #1,#2 \Rightarrow #3}
\newcommand{\gexec}[3]{ #1,#2 \mathrel{\Rightarrow_{\mathrm{g}}} #3}

\newcommand{\execseq}[3]{#1,#2 \Rightarrow^\mathrm{seq} #3}
\newcommand{\gexecseq}[3]{#1,#2 \mathrel{\Rightarrow^\mathrm{seq}_{\mathrm{g}}} #3}
\newcommand{\execparR}[3]{#1,#2 \Rightarrow^\mathrm{parR} #3}
\newcommand{\gexecparR}[3]{#1,#2 \mathrel{\gg^\mathrm{R}_\mathrm{g}} #3}
\newcommand{\execparL}[3]{#1,#2 \Rightarrow^\mathrm{parL} #3}
\newcommand{\gexecparL}[3]{#1,#2 \mathrel{\gg^\mathrm{L}_\mathrm{g}} #3}

\newcommand{\atomize}[2]{#1 \leadsto #2}
\newcommand{\gatomize}[2]{#1 \mathrel{\leadsto_\mathrm{g}} #2}

\newcommand{\st}{\sigma}
\newcommand{\state}{\mathit{state}}

\newcommand{\sem}[2]{\llbracket #1 \rrbracket~#2}

\newcommand{\eval}{\mathit{eval}}
\newcommand{\evalseq}{\mathit{evalseq}}

\newcommand{\evalparR}{\mathit{evalparR}}
\newcommand{\evalparL}{\mathit{evalparL}}
\newcommand{\close}{\mathit{close}}

\newcommand{\geval}{\mathit{eval}_{\mathrm{g}}}
\newcommand{\gevalseq}{\mathit{evalseq}_{\mathrm{g}}}

\newcommand{\gmergeR}{\mathit{mergeR}_{\mathrm{g}}}
\newcommand{\gmergeL}{\mathit{mergeL}_{\mathrm{g}}}
\newcommand{\gclose}{\mathit{close}_{\mathrm{g}}}

\newcommand{\dn}{\mathrel{\downarrow}}
\newcommand{\divg}{\uparrow}

\newcommand{\trace}{\mathit{trace}}
\newcommand{\gtrace}{\mathit{trace}_\mathrm{g}}

\begin{document}

\title{Coinductive Big-Step Semantics for Concurrency}

\def\titlerunning{Coinductive Big-Step Semantics for Concurrency}

\author{
Tarmo Uustalu
\institute{Institute of Cybernetics at Tallinn University of Technology, %\\
Akadeemia tee 21, 12618 Tallinn, Estonia} 
\email{tarmo@cs.ioc.ee}
}

\def\authorrunning{T.~Uustalu}

\maketitle

\begin{abstract}
  In a paper presented at SOS 2010 \cite{NU:reswbb}, we developed a
  framework for big-step semantics for interactive input-output in
  combination with divergence, based on coinductive and mixed
  inductive-coinductive notions of resumptions, evaluation and
  termination-sensitive weak bisimilarity. In contrast to standard
  inductively defined big-step semantics, this framework handles
  divergence properly; in particular, runs that produce some
  observable effects and then diverge, are not ``lost''. Here we scale
  this approach for shared-variable concurrency on a simple example
  language. We develop the metatheory of our semantics in a
  constructive logic.
\end{abstract}

\section{Introduction}

The purpose of this paper is to advocate two ideas. First, big-step
operational semantics can handle divergence as well as small-step
semantics, so that both terminating and diverging behaviors can be
reasoned about uniformly. Big-step semantics that account for
divergence properly are achieved by working with coinductive semantic
entities (transcripts of possible infinite computation paths or
nonwellfounded computation trees) and coinductive evaluation. Second,
contrary to what is so often stated, concurrency is not inherently
small-step, or at least not more inherently than any kind of effect
produced incrementally during a program's run (e.g., interactive
output). Big-step semantics for concurrency can be built by borrowing
the suitable denotational machinery, except that we do not want to use
domains and fixpoints to deal with partiality, but coinductively
defined sets and corecursion. In this paper, we use resumptions, more
specifically coinductive resumptions.

We build on our previous work~\cite{NU:reswbb} and develop two
resumption-based big-step semantics for a simple imperative language
with shared-variable concurrency. The metatheory of these
semantics---e.g., the equivalence of evaluation in the big-step
semantics to maximal multi-step reduction in a reference small-step
semantics---is entirely constructive---meaning that we can compute
evaluations from maximal multi-step reductions and vice
versa. Moreover, evaluation is deterministic and can be turned into a
computable function.

The idea that divergence can be properly accounted for by switching to
coinductively defined semantic entities such as possibly infinitely
delayed states or possibly infinite traces is due to Capretta
\cite{Cap:genrct}. The deeper underlying theory is based on completely
iterative monads and has been treated in detail by Goncharov and
Schr\"oder~\cite{GS:coicas}. 

Leroy and Grall~\cite{LG:coibso} attempted to use coinductive big-step
semantics to reason about both terminating and diverging program runs
in the Compcert project on a formally certified compiler, but ran into
certain semantic anomalies (proving the big-step and small-step
semantics equivalent required the use of excluded middle, which should
not be needed; infinite loops were not specifically arranged to be
productive, with the effect that infinite loops with no observable
effects led to finite traces, to which other traces could be appended).
(Cf. also the simultaneous work by Cousot and Cousot \cite{CC:biios}.)
Nakata and Uustalu \cite{NU:trabco} fixed the anomalies and arrived at
a systematic account of trace-based big-step semantics for divergence
in a purely sequential, side-effect-free setting (in relational and
also functional styles). Further \cite{NU:hoalct,NU:reswbb}, they also
developed a matching Hoare logic and a resumption-based big-step
semantics for a combination of interactive input/output with
divergence.  Danielsson~\cite{Dan:opespm} has promoted especially
functional-style coinductive big-step semantics.
Ancona~\cite{Anc:souool} used a coinductive big-step semantics of Java
to show it type-sound in a sense that covers also divergence: if a
program is type-sound, it produces a trace.

The tool of resumptions was originated by Plotkin~\cite{Plo:powc} and
has since been developed and used by several authors
\cite{CM:synmds,Har:essmt}.
An inductive trace-based big-step semantics for a concurrent language
(not handling divergence) has appeared in the work of
Mitchell~\cite{Mit:conns}.

The paper is organized as follows. In Section~\ref{sec:sem}, we
introduce our example language with pre-emptive scheduling, give it
two resumption-based big-step semantics and relate them to a
small-step semantics. In Section~\ref{sec:equiv}, we discuss notions
of equivalence of resumptions. In Section~\ref{sec:funct}, we show
that our semantics can also be formulated functionally rather than
relationally. In Section~\ref{sec:traces}, we discuss the major
alternative to resumptions---traces. We conclude in
Section~\ref{sec:concl}.

In Appendix~\ref{sec:coop}, we consider cooperative scheduling.

Haskell implementations of the functional-style semantics of
Section~\ref{sec:funct} are available online at
\url{http://cs.ioc.ee/~tarmo/papers/}, to be completed with an Agda
formalization of the whole paper.

\section{An  example language and resumption-based semantics}
\label{sec:sem}

\subsection{Syntax}
\label{sec:syntax}

We look at a minimal language with shared-variable concurrency
(cf.~Amadio~\cite{Ama:opemc}) whose statements are given inductively
by the grammar
\[
s  ::= 
 \Assign{x}{e} \mid \Skip \mid \Seq{s_0}{s_1} \mid \Ifthenelse{e}{s_t}{s_f}
\mid \While{e}{s_t} 
\mid s_0 \parallel s_1 \mid \Atomic{s} \mid \Await{e}{s}
\]

The intention is that $s_0 \parallel s_1$ is parallel composition of
$s_0$ and $s_1$ (in particular, it terminates when both branches have
terminated). The statement $\Atomic{s}$ is executed by running $s$
atomically; the statement $\Await{e}{s}$ is executed by waiting until
$e$ is true (other computations can have their chance in the meantime)
and then running $s$ atomically. Throughout the paper proper,
scheduling is preemptive, with only assignments and boolean guards
atomic implicitly.

In Appendix~\ref{sec:coop}, we look at a cooperative scheduling
interpretation of the same syntax.

\subsection{Big-step semantics}
\label{sec:big}

We will first introduce a semantics that captures all runs of a
statement from a state until the closest control release points in one
step. (In the next section, we will introduce a semantics that deals
with return of control.)

The central semantic entities of this semantics are
\emph{resumptions} (computation trees). Resumptions are defined
coinductively by the following rules (in this text, inductive
definitions are shown by single rule lines, coinductive definitions
are indicated by double rule-lines).
\[
\infer={ \ret{\st} : \res}{ \st : \state}
\quad
\infer={
  \delay{r} : \res
}{
  r : res
}
\quad
\infer={
  r_0 + r_1 : \res
}{
  r_0 : \res
  & 
  r_1 : \res
}
\quad
\infer={
  \yield{s}~\st : \res
}{
  s : \stmt
  & 
  \st : \state
}
\]
The resumption $\ret{\st}$ denotes a computation that terminated in a
state $\st$. The resumption $\delay{r}$ is a computation that first
produces an unit delay (makes an internal small step) and continues
then as $r$. The resumption $r_0 + r_1$ is a choice between two
resumptions $r_0$ and $r_1$. The resumption $\yield{s}~\st$ is a
computation that has released control in a state $\st$ and will
further execute a statement $s$ when (and if) it regains control
(notice the presence of a syntactic entity here!). The definition
being coinductive has the effect that resumptions can be
non-wellfounded, i.e., computations can go on forever.

E.g., the following is a resumption that involves some internal small
steps, two choices; one path terminates, one diverges, one suspends:
\[
\delta^3\, (\delta^2\, (\ret{[x \mapsto 5]}) 
 + \delta^4\, (\delta^\infty + \delta\, (\yield{x := x+7}~[x \mapsto 3]))
\]
(by $\delta^\infty$ we mean the diverging resumption defined
corecursively by $\delta^\infty = \delta~ \delta^\infty$).

\emph{Evaluation} of a statement $s$ relates a (pre-)state to a
(post-)resumption and is defined coinductively by the rules
\[
\small
\begin{array}{c}
\infer={\exec{\Assign{x}{e}}{\st}{\delay{(\ret{\update{\st}{x}{\sem{e}{\st}}})}
}
}{}
\qquad
\infer={\exec{\Skip}{\st}{\ret{\st}}
}{}
\quad
\infer={
  \exec{\Seq{s_0}{s_1}}{\st}{r'}
}{
  \exec{s_0}{\st}{r}
  &\execseq{s_1}{r}{r'}
}
\\[1ex]
\infer={
  \exec{\Ifthenelse{e}{s_t}{s_f}}{\st}{\delay{(\yield{s_t}~\st})}
}{
  \istrue{\st}{e}
}
\quad
\infer={
  \exec{\Ifthenelse{e}{s_t}{s_f}}{\st}{\delay{(\yield{s_f}~\st})}
}{
  \isfalse{\st}{e}
}
\\[1ex]
\infer={
  \exec{\While{e}{s_t}}{\st}{\delay{(\yield{(s_t; \While{e}{s_t})}~\st)}}
}{
  \istrue{\st}{e}
}
\quad
\infer={
  \exec{\While{e}{s_t}}{\st}{\delay{(\ret{\st})}}
}{
  \isfalse{\st}{e}
}
\\[1ex]
\infer={
  \exec{s_0 \parallel s_1}{\st}{r_0' + r_1'}
}{
  \exec{s_0}{\st}{r_0}
  &
  \execparR{s_1}{r_0}{r_0'}
  &
  \exec{s_1}{\st}{r_1}
  &
  \execparL{s_0}{r_1}{r_1'}
}
\quad
\infer={
  \exec{\Atomic{s}}{\st}{r'}
}{
  \exec{s}{\st}{r}
  &
  \atomize{r}{r'}
}
\\[1ex]
\infer={
  \exec{\Await{e}{s}}{\st}{\delay{r'}}
}{
  \st \models e
  &
  \exec{s}{\st}{r}
  &
  \atomize{r}{r'}
}
\quad
\infer={
  \exec{\Await{e}{s}}{\st}{\delay{(\yield{(\Await{e}{s})}~\st)}}
}{
  \st \not\models e
}
\end{array}
\]
We have made sure that internal small steps take their time by
inserting unit delays at all places where assignments or boolean
guards are evaluated. This makes evaluation deterministic, allowing us
to turn it into a function, as we will see later in
Sec.~\ref{sec:funct}. The $\mathit{yield}$s in the rules for
$\textsf{await}$, $\textsf{if}$ and $\textsf{while}$ signify control
release points.  Control release also occurs at the ``midpoint'' of
evaluation of any sequential or parallel composition (i.e., at the
termination of the first resp.\ faster statement). This is handled by
the $\mathit{ret}$ rules for sequential and parallel extensions of
evaluation.

\emph{Sequential extension of evaluation} relates a (pre-)resumption
(the resumption present before some statement is evaluated) to
a (post-)resumption (the total resumption after). It is defined
coinductively by the rules
\[
\small
\infer={
  \execseq{s}{\ret{\st}}{\yield{s}~\st}
}{
}
\quad
\infer={
  \execseq{s}{\delay{r}}{\delay{r'}}
}{
  \execseq{s}{r}{r'}
}
\quad
\infer={
  \execseq{s}{r_0 + r_1}{r_0' + r_1'}
}{
  \execseq{s}{r_0}{r_0'}
  &
  \execseq{s}{r_1}{r_1'}
}
\quad
\infer={
  \execseq{s}{\yield{s_0}~\st}{\yield{(s_0 ; s)}~\st}
}{
}
\]
Essentially, sequential extension of evaluation is a form of
coinductive prefix closure of evaluation. But, in addition, the
$\mathit{ret}$ rule inserts a control release between the termination
of the first statement and the start of the second statement of a
sequential composition. In the case of a $\mathit{yield}$
pre-resumption, we simply grow the residual statement.

\emph{Parallel extension of evaluation}, which also relates a
resumption to a resumption, is for evaluating a given statement in
parallel with a given resumption. The idea is to create an opportunity
for the given statement to start when (and if) the resumption
terminates or releases control.  Also this relation is defined
coinductively. Also here, in the base case (where the given resumption
has terminated), we have a control release point.
\[
\small
\begin{array}{c}
\infer={
  \execparR{s}{\ret{\st}}{\yield{s}~\st}
}{
}
\quad
\infer={
  \execparR{s}{\delay{r}}{\delay{r'}}
}{
  \execparR{s}{r}{r'}
}
\quad
\infer={
  \execparR{s}{r_0 + r_1}{r_0' + r_1'}
}{
  \execparR{s}{r_0}{r_0'}
  &
  \execparR{s}{r_1}{r_1'}
}
\quad
\infer={
  \execparR{s}{\yield{s_0}~\st}{\yield{(s_0 \parallel s)}~\st}
}{
}
\\[1ex]
\infer={
  \execparL{s}{\ret{\st}}{\yield{s}~\st}
}{
}
\quad
\infer={
  \execparL{s}{\delay{r}}{\delay{r'}}
}{
  \execparL{s}{r}{r'}
}
\quad
\infer={
  \execparL{s}{r_0 + r_1}{r_0' + r_1'}
}{
  \execparL{s}{r_0}{r_0'}
  &
  \execparL{s}{r_1}{r_1'}
}
\quad
\infer={
  \execparL{s}{\yield{s_1}~\st}{\yield{(s \parallel s_1)}~\st}
}{
}
\end{array}
\]

Finally, \emph{closing a resumption} makes sure it does not release
control.  This is done by (repeatedly) ``stitching'' a resumption at
every control release point by evaluating the residual statement from
the state at this point. The corresponding relation between two
resumptions is defined coinductively by
\[
\small
\infer={
  \atomize{\ret{\st}}{\ret{\st}}
}{
}
\quad
\infer={
  \atomize{\delay{r}}{\delay{r'}}
}{
  \atomize{r}{r'}
}
\quad
\infer={
  \atomize{r_0 + r_1}{r_0' + r_1'}
}{
  \atomize{r_0}{r_0'}
  &
  \atomize{r_1}{r_1'}
}
\quad
\infer={
  \atomize{\yield{s}~\st}{\delay{r'}}
}{
  \exec{s}{\st}{r}
  &
  \atomize{r}{r'}
}
\]
In the last rule, the constructor $\mathit{yield}$ does not disappear
without leaving a trace, it is replaced with $\mathit{delay}$,
corresponding to an internal small step. 

To give only two smallest examples, for $s = x := 1 \parallel (x :=
x+2; x := x+2)$, $\st = [x \mapsto 0]$, we have
\[
\exec{s}{\st}{\delta(\yield{(x := x+2; x := x+2)}~[x \mapsto 1]) +
  \delta(\yield{(x :=1 \parallel x := x+2)}~[x \mapsto 2])}
\]
while
\[
\exec{\Atomic{s}}{\st}{\delta^5{(\ret{[x \mapsto 5])}} + \delta^2{(\delta^3{(\ret{[x \mapsto 3]})}+\delta^3{(\ret{[x \mapsto 1]})})}}
\]
For $s = (\Await{x = 0}{x :=1})  \parallel x :=2$, $\st = [x \mapsto 0]$,
we have 
\[
\exec{s}{\st}{\delta^2{(\yield{x :=2}~[x \mapsto 1])} + \delta^1{(\yield{(\Await{x = 0}{x :=1})}~[x \mapsto 2])}}
\]
whereas 
\[
\exec{\Atomic{s}}{\st}{\delta^4{(\ret [x \mapsto 2])} + \delta^\infty}
\]

In this semantics there is no fairness, all schedules are considered.
The resumption for statement $\Atomic{(x := 1
  \parallel \While{x = 0}{\Skip})}$ and state $[x \mapsto 0]$
contains a path that never terminates. Note that fairness is a
property of a path in a resumption, not of a resumption. Being
an inductive property, fairness cannot be refuted based on an initial
segment of a path, so unfair paths cannot be cut out of a resumption.

\subsection{Giant-step semantics}
\label{sec:giant}

An alternative to what we have considered in the previous section is
to run statements beyond control release points for any states that
control may potentially be returned in, i.e., for all states.

This leads to what we call a giant-step semantics here in order to
have a different name for it.\footnote{One might, of course, argue,
  that what I have called the ``big-step'' semantics here should be
  called ``medium-step'', and the ``giant-step'' semantics should be
  called ``big-step''.  I would not disagree at all. My choice of
  terminology here was motivated by the intuition that ``big-step''
  evaluation should run a statement to its completion.  When a
  statement's run has reached a control release point, it is complete
  in the sense that it cannot run further on its own; what will happen
  further depends on the scheduler (it might even be unfair and not
  return control to it at all).  Note, however, that big-step and
  giant-step evaluation agree fully for statements of the form
  $\Atomic{s}$.}

In this semantics, resumptions are purely semantic, they do not
contain any statement syntax. They are defined as before, except that
the $\mathit{yield}$ constructor is typed differently.
\[
\infer={ \ret{\st} : \gres}{ \st : \state}
\quad
\infer={
  \delay{r} : \gres
}{
  r : res
}
\quad
\infer={
  r_0 + r_1 : \gres
}{
  r_0 : \gres
  & 
  r_1 : \gres
}
\quad
\infer={
  \yield{k}~\st : \gres
}{
  k : \state \to \gres
  & 
  \st : \state
}
\]
$\yield{k}~\st$ is a resumption that has released control in a given
state $\st$ and, when returned control in some state $\st'$, will
continue as $k~\st'$. We call functions from states to resumptions
\emph{continuations}.

Evaluation is defined essentially as before, but with appropriate
adjustments, as what were residual statements must now be evaluated.
\[
\small
\begin{array}{c}
\infer={\gexec{\Assign{x}{e}}{\st}{\delay{(\ret{\update{\st}{x}{\sem{e}{\st}}})}
}
}{}
\\[1ex]
\infer={\gexec{\Skip}{\st}{\ret{\st}}
}{}
\quad
\infer={
  \gexec{\Seq{s_0}{s_1}}{\st}{r'}
}{
  \gexec{s_0}{\st}{r}
  &\gexecseq{s_1}{r}{r'}
}
\\[1ex]
\infer={
  \gexec{\Ifthenelse{e}{s_t}{s_f}}{\st}{\delay{(\yield{k}~\st})}
}{
  \istrue{\st}{e}
  &
  \forall \st'.\, \gexec{s_t}{\st'}{k~ \st'}
}
\quad
\infer={
  \gexec{\Ifthenelse{e}{s_t}{s_f}}{\st}{\delay{(\yield{k}~\st})}
}{
  \isfalse{\st}{e}
  &
  \forall \st'.\, \gexec{s_f}{\st'}{k ~ \st'}
}
\\[1ex]
\infer={
  \gexec{\While{e}{s_t}}{\st}{\delay{(\yield{k'}~\st)}}
}{
  \istrue{\st}{e}
  &
  \forall \st'.\,\, \gexec{s_t}{\st'}{k~ \st'}
  &
  \forall \st'.\,\, \gexecseq{\While{e}{s_t}}{k~ \st'}{k'~ \st'} 
}
\quad
\infer={
  \gexec{\While{e}{s_t}}{\st}{\delay{(\ret{\st})}}
}{
  \isfalse{\st}{e}
}
\\[1ex]
\infer={
  \gexec{s_0 \parallel s_1}{\st}{r_0' + r_1'}
}{
  \gexec{s_0}{\st}{r_0}
  &
  \forall \st'. \gexec{s_1}{\st'}{k_1~\st'}
  &
  \gexecparR{k_1}{r_0}{r_0'}
  &
  \gexec{s_1}{\st}{r_1}
  &
  \forall \st'. \gexec{s_0}{\st'}{k_0~\st'} 
  &
  \gexecparL{k_0}{r_1}{r_1'}
}
\\[1ex]
\infer={
  \gexec{\Atomic{s}}{\st}{r'}
}{
  \gexec{s}{\st}{r}
  &
  \gatomize{r}{r'}
}
\\[1ex]
\infer={
  \gexec{\Await{e}{s}}{\st}{\delay{r'}}
}{
  \istrue{\st}{e}
  &
  \gexec{s}{\st}{r}
  &
  \gatomize{r}{r'}
}
\quad
\infer={
  \gexec{\Await{e}{s}}{\st}{\delay{(\yield{k}~\st)}}
}{
  \isfalse{\st}{e}
  &
  \forall \st'.\,\, \gexec{\Await{e}{s}}{\st'}{k~ \st'}
}
\end{array}
\]

In sequential extension of evaluation, the rule for $\mathit{yield}$
is now similar to those for $\delta$ and $+$, so we are dealing with a
proper coinductive prefix closure of the evaluation relation modulo
the extra $\mathit{yield}$ constructor in the $\mathit{ret}$ rule to
cater for control release at the midpoint of evaluation of a
sequential composition.
\[
\small
\infer={
  \gexecseq{s}{\ret{\st}}{\yield{k}~\st}
}{
  \forall \st'.\,\, \gexec{s}{\st'}{k\, \st'}
}
\quad
\infer={
  \gexecseq{s}{\delay{r}}{\delay{r'}}
}{
  \gexecseq{s}{r}{r'}
}
\quad
\infer={
  \gexecseq{s}{r_0 + r_1}{r_0' + r_1'}
}{
  \gexecseq{s}{r_0}{r_0'}
  &
  \gexecseq{s}{r_1}{r_1'}
}
\quad
\infer={
  \gexecseq{s}{\yield{k}~\st}{\yield{k'}~\st}
}{
  \forall \st'.\,\, \gexecseq{s}{k~ \st'}{k'\, \st'}  
}
\]

Instead of parallel extension of evaluation, we define \emph{merging a
  continuation into a resumption}.
\[
\small
\begin{array}{c}
\infer={
  \gexecparR{k}{\ret{\st}}{\yield{k}~\st}
}{
}
\quad
\infer={
  \gexecparR{k}{\delay{r}}{\delay{r'}}
}{
  \gexecparR{k}{r}{r'}
}
\quad
\infer={
  \gexecparR{k}{r_0 + r_1}{r_0' + r_1'}
}{
  \gexecparR{k}{r_0}{r_0'}
  &
  \gexecparR{k}{r_1}{r_1'}
}
\\[10pt]
\infer={
  \gexecparR{k}{\yield{k_0}~\st}{\yield{(\lambda \st'.\, k_0'~\st' + k_1'~\st')}~\st}
}{
  \forall \st'.\,\, \gexecparR{k}{k_0~\st'}{k_0'~\st'}
  &
  \forall \st'.\,\, \gexecparL{k_0}{k~\st'}{k_1'~\st'}
}\\[1ex]
\infer={
  \gexecparL{k}{\ret{\st}}{\yield{k}~\st}
}{
}
\quad
\infer={
  \gexecparL{k}{\delay{r}}{\delay{r'}}
}{
  \gexecparL{k}{r}{r'}
}
\quad
\infer={
  \gexecparL{k}{r_0 + r_1}{r_0' + r_1'}
}{
  \gexecparL{k}{r_0}{r_0'}
  &
  \gexecparL{k}{r_1}{r_1'}
}
\\[10pt]
\infer={
  \gexecparL{k}{\yield{k_1}~\st}{\yield{(\lambda \st'.\, k_0'~\st' + k_1'~\st')}~\st}
}{
  \forall \st'.\,\, \gexecparR{k_1}{k~\st'}{k_0'~\st'}
  &
  \forall \st'.\,\, \gexecparL{k}{k_1~\st'}{k_1'~\st'}
}
\end{array}
\]
Here, in the rules for $\mathit{yield}$, we construct continuations
corresponding to evaluating suitable $\parallel$ statements from any
given states.

Closing a resumption is straightforward. To close a $\mathit{yield}$
resumption, we apply the given continuation to the given state, close
the resulting resumption and add a unit delay.
\[
\small
\infer={
  \gatomize{\ret{\st}}{\ret{\st}}
}{
}
\quad
\infer={
  \gatomize{\delay{r}}{\delay{r'}}
}{
  \gatomize{r}{r'}
}
\quad
\infer={
  \gatomize{r_0 + r_1}{r_0' + r_1'}
}{
  \gatomize{r_0}{r_0'}
  &
  \gatomize{r_1}{r_1'}
}
\quad
\infer={
  \gatomize{\yield{k}~\st}{\delay{r}}
}{
  \gatomize{k\, \st}{r}
}
\]

For example, $s = x := 1 \parallel (x := x+2; x := x+2)$, $\st = [x \mapsto
0]$, we have
\[
\begin{array}{l}
\gexec{s}{\st}{}\delta(\yield{(\lambda \st'.\, \delta(\yield{(\lambda \st''.\, \delta(\ret{\st''[x \mapsto \st''~ x+ 2]}))}~[x \mapsto \st'~x + 2]))}~[x \mapsto 1]) \\
\hspace*{10mm} +
  \delta(\yield{(\lambda \st'.\, \delta(\yield{(\lambda \st''.\, \delta(\ret{\st''[x \mapsto \st''~x + 2]}))}~[x \mapsto 1]) \\
\hspace*{32mm}
            + \delta(\yield{(\lambda \st''.\, \delta(\ret{[x \mapsto 1]}))}~[x \mapsto \st'~x + 2]))}~ [x \mapsto 2])
\end{array}
\]

\subsection{Small-step semantics}
\label{sec:small}

To validate the big-step semantics of Sec.~\ref{sec:big}, we can
compare it to a small-step semantics.

To get a close match with the big-step semantics, where we capture all
runs of a program in a single resumption, we give a (perhaps somewhat
nonstandard) small-step semantics that makes it possible to keep track
of all runs of a statement at once.

This semantics works with \emph{extended configurations}. They are
defined as follows. (We use the notation of an inductive definition,
but in fact this datatype is a simple disjoint union.)
\[
\infer{ \ret{\st} : \xcfg}{ \st : \state}
\quad
\infer{
  \delay{(s, \st)} : \xcfg
}{
  s : \stmt
  &
  \st : \state
}
\quad
\infer{
  (s_0, \st_0) + (s_1, \st_1) : \xcfg
}{
  s_0 : \stmt
  &
  \st_0 : \state
  & 
  s_1 : \stmt
  &
  \st_1 : \state
}
\quad
\infer{
  \yield{s}~\st : \xcfg
}{
  s : \stmt
  & 
  \st : \state
}
\]
$\ret{\st}$ is a terminated computation. $\delay{(s, \st)}$ is a
computation that after an internal step is in a state $\st$ and has
$s$ to execute yet. $(s_0, \st_0) + (s_1, \st_1)$ is a computation
that makes a choice and is then in a state $\st_0$ with $s_0$ to
execute or in a state $\st_1$ with $s_1$ to execute. $\yield{s}~\st$
is a computation that has released control in a state $\st$ and has
$s$ to execute when (and if) it regains control.

\emph{Reduction} relates a state to an extended configuration and is
defined inductively(!). So small steps are justified by finite
derivations.
\[
\small
\begin{array}{c}
\infer{\red{x:=e}{\st}{\delay{(\Skip, \st[x \mapsto \sem{e} \st])}}
}{
}
\\[1ex]
\infer{\red{\Skip}{\st}{\ret{\st}}
}{
}
\\[1ex]
\infer{\red{s_0; s_1}{\st}{\yield{s_1}{\st'}}
}{
  \red{s_0}{\st}{\ret{\st'}}
}
\quad
\infer{\red{s_0; s_1}{\st}{\delay{(s_0' ; s_1, \st')}}
}{
  \red{s_0}{\st}{\delay{(s_0',\st')}}
}
\\[1ex]
\infer{\red{s_0; s_1}{\st}{(s_{00} ; s_1,\st_0) + (s_{01}; s_1,\st_1)}
}{
  \red{s_0}{\st}{(s_{00},\st_0) + (s_{01},\st_1)}
}
\quad
\infer{\red{s_0; s_1}{\st}{\yield{(s_0' ; s_1)}~ \st'}
}{
  \red{s_0}{\st}{\yield{s_0'}~\st'}
}
\\[1ex]
\infer{\red{\Ifthenelse{e}{s_t}{s_f}}{\st}{\delay{(\Skip; s_t, \st)}}
}{
  \st \models e
}
\quad
\infer{\red{\Ifthenelse{e}{s_t}{s_f}}{\st}{\delay{(\Skip; s_f, \st)}}
}{
 \st \not\models e
}
\\[1ex]
\infer{\red{\While{e}{s_t}}{\st}{\delay{(\Skip; (s_t ; \While{e}{s_t}), \st)}}
}{
 \st \models e
}
\quad
\infer{\red{\While{e}{s_t}}{\st}{\delay{(\Skip, \st)}}
}{
 \st \not\models e
}
\end{array}
\]
\[
\small
\begin{array}{c}
\infer{\red{s_0 \parallel s_1}{\st}{(s_0 \lpar s_1, \st) + (s_0 \rpar s_1, \st)}
}{
}
\\[1ex]
\infer{\red{s_0 \lpar s_1}{\st}{\yield{s_1}{\st'}}
}{
  \red{s_0}{\st}{\ret{\st'}}
}
\quad
\infer{\red{s_0 \lpar s_1}{\st}{\delay{(s_0' \lpar s_1, \st')}}
}{
  \red{s_0}{\st}{\delay{(s_0',\st')}}
}
\\[1ex]
\infer{\red{s_0 \lpar s_1}{\st}{(s_{00} \lpar s_1,\st_0) + (s_{01} \lpar s_1,\st_1)}
}{
  \red{s_0}{\st}{(s_{00},\st_0) + (s_{01},\st_1)}
}
\quad
\infer{\red{s_0 \lpar s_1}{\st}{\yield{(s_0' \parallel s_1)}~ \st'}
}{
  \red{s_0}{\st}{\yield{s_0'}~\st'}
}
\\[1ex]
\infer{\red{s_0 \rpar s_1}{\st}{\yield{s_0}{\st'}}
}{
  \red{s_1}{\st}{\ret{\st'}}
}
\quad
\infer{\red{s_0 \rpar s_1}{\st}{\delay{(s_0 \rpar s_1', \st')}}
}{
  \red{s_1}{\st}{\delay{(s_1',\st')}}
}
\\[1ex]
\infer{\red{s_0 \rpar s_1}{\st}{(s_0 \rpar s_1,\st_{10}) + (s_0 \rpar s_1,\st_{11})}
}{
  \red{s_1}{\st}{(s_{10},\st_0) + (s_{11},\st_1)}
}
\quad
\infer{\red{s_0 \rpar s_1}{\st}{\yield{(s_0 \parallel s_1')}~ \st'}
}{
  \red{s_1}{\st}{\yield{s_1'}~\st'}
}
\\[1ex]
\infer{\red{\Atomic{s}}{\st}{\ret{\st'}}
}{
  \red{s_}{\st}{\ret{\st'}}
}
\quad
\infer{\red{\Atomic{s}}{\st}{\delay{(\Atomic{s'}, \st')}}
}{
  \red{s}{\st}{\delay{(s',\st')}}
}
\\[1ex]
\infer{\red{\Atomic{s}}{\st}{(\Atomic{s_0},\st_0) + (\Atomic{s_1},\st_1)}
}{
   \red{s}{\st}{(s_0,\st_0) + (s_1,\st_1)}
}
\quad
\infer{\red{\Atomic{s}}{\st}{\delay{(\Atomic{s'}, \st')}}
}{
  \red{s}{\st}{\yield{s'}~\st'}
}
\\[1ex]
\infer{\red{\Await{e}{s}}{\st}{\delay{(\Atomic{s}, \st)}}
}{
 \st \models e
}
\quad
\infer{\red{\Await{e}{s}}{\st}{\delay{(\Skip; \Await{e}{s}, \st)}}
}{
 \st \not\models e
}
\end{array}
\]
Notice that $\Skip; s$ differs from $s$ by allowing a control release
before $s$ is started. To use some device like this is unavoidable, if
we want the reduction relation to capture exactly one small step
leading to a configuration. We have also used auxiliary statement
forms $s_0 \lpar s_1$ and $s_0 \rpar s_1$ which are like parallel
composition except that $s_0$ resp.\ $s_1$ makes the first small step.

\emph{Maximal multi-step reduction} relates a state to a resumption
and is defined coinductively:
\[
\small
\begin{array}{c}
\infer={\mmred{s}{\st}{\ret{\st'}}
}{
  \red{s}{\st}{\ret{\st'}}
}
\quad
\infer={\mmred{s}{\st}{\delay{r}}
}{
  \red{s}{\st}{\delay{(s',\st')}}
  &
  \mmred{s'}{\st'}{r}
}
\\[1ex]
\infer={\mmred{s}{\st}{r_0 + r_1}
}{
  \red{s}{\st}{(s_0,\st_0) + (s_1, \st_1)}
  &
  \mmred{s_0}{\st_0}{r_0}
  &
  \mmred{s_1}{\st_1}{r_1}
}
\quad
\infer={\mmred{s}{\st}{\yield{s'}~\st'}
}{
  \red{s}{\st}{\yield{s'}~\st'}
}
\end{array}
\]
It applies single-step reduction repeatedly as many times as possible
viewing $\mathit{ret}$ and $\mathit{yield}$ configurations as terminal
and develops a resumption.

Evaluation of the big-step semantics agrees with maximal multi-step
reduction: $\exec{s}{\st}{r}$ iff $\mmred{s}{\st}{r}$.

A variation of maximal multi-step reduction that also reduces under
$\mathit{yield}$s develops a resumption of the giant-step semantics of
Sec.~\ref{sec:giant}.
\[
\small
\begin{array}{c}
\infer={\gmmred{s}{\st}{\ret{\st'}}
}{
  \red{s}{\st}{\ret{\st'}}
}
\quad
\infer={\gmmred{s}{\st}{\delay{r}}
}{
  \red{s}{\st}{\delay{(s',\st')}}
  &
  \gmmred{s'}{\st'}{r}
}
\\[1ex]
\infer={\gmmred{s}{\st}{r_0 + r_1}
}{
  \red{s}{\st}{(s_0,\st_0) + (s_1, \st_1)}
  &
  \gmmred{s_0}{\st_0}{r_0}
  &
  \gmmred{s_1}{\st_1}{r_1}
}
\quad
\infer={\gmmred{s}{\st}{\yield{k}~\st'}
}{
  \red{s}{\st}{\yield{s'}~\st'}
  &
  \forall \st''.\, \gmmred{s'}{\st''}{k~\st''}
}
\end{array}
\]

Evaluation of the giant-step semantics agrees with this variation of
maximal multi-step reduction: $\gexec{s}{\st}{r}$ iff
$\gmmred{s}{\st}{r}$.

\section{Equivalences of resumptions}
\label{sec:equiv}

When are two resumptions to be considered equivalent? This depends on
the purpose at hand. The finest sensible notion is \emph{strong
  bisimilarity} defined for big-step resumptions coinductively by the
rules
\[
\infer={
  \bism{\ret{\st}}{\ret{\st}}
}{}
\quad
\infer={
  \bism{\delay{r}}{\delay{r_*}}
}{
  \bism{r}{r_*}
}
\quad
\infer={
  \bism{r_0 + r_1}{r_{0*} + r_{1*}}
}{
  \bism{r_0}{r_{0*}}
  &
  \bism{r_1}{r_{1*}}
}
\quad
\infer={
  \bism{\yield{s}~\st}{\yield{s}~\st}
}{
}
\]
Classically, this predicate is just equality of big-step resumptions.
But in intensional type theory, propositional equality is stronger;
strong bisimilarity as just defined does not imply propositional
equality\footnote{This phenomenon is similar to extensional function
  equality, i.e., propositional equality of two functions on all
  arguments: it does not imply propositional equality of the
  functions.}.

Strong bisimilarity in the sense just defined may feel entirely
uninteresting.  Yet it is meaningful and important constructively.
E.g., big-step evaluation is deterministic up to strong bisimilarity,
but not up to propositional equality.

For giant-step resumptions, strong bisimilarity is defined by the
rules
\[
\infer={
  \gbism{\ret{\st}}{\ret{\st}}
}{}
\quad
\infer={
  \gbism{\delay{r}}{\delay{r_*}}
}{
  \gbism{r}{r_*}
}
\quad
\infer={
  \gbism{r_0 + r_1}{r_{0*} + r_{1*}}
}{
  \gbism{r_0}{r_{0*}}
  &
  \gbism{r_1}{r_{1*}}
}
\quad
\infer={
  \gbism{\yield{k}~\st}{\yield{k_*}~\st}
}{
  \forall \st'.\, \gbism{k~\st'}{k_*~\st'}
}
\]

The useful coarser notions ignore order and multiplicity of choices
(strong bisimilarity as in process algebras), exact durations of
finite delays (termination-sensitive weak bisimilarity) or both.  The
definition of termination-sensitive weak bisimilarity requires
combining or mixing induction and coinduction, with several caveats to
avoid. First, it is easy to misdefine weak bisimilarity so that it
equates any resumption with the divergent resumption and therefore all
resumptions. Second, a fairly attractive definition fails to give
reflexivity without the use of excluded middle, which is a warning
that the definition is not the ``right one'' from the constructive
point of view.

Let us look at the definition of weak bisimilarity for big-step
resumptions. First we define \emph{convergence} of a resumption
inductively by the rules
\[
\begin{array}{c}
\infer{\ret{\st} \dn \ret{\st}}{
}
\quad 
\infer{\delay{r} \dn r'}{
  r \dn r'
}
\quad 
\infer{r_0 + r_1 \dn r_0' + r_1'}{
  r_0 \dn r_0'
  &
  r_1 \dn r_1'
}
\quad
\infer{\yield{s}~\st \dn \yield{s}~\st}{
}
\end{array}
\]
Intuitively, a resumption $r$ converges to another resumption $r'$ if
$r'$ is a $\mathit{ret}$, $+$ or $\mathit{yield}$ resumption and can
be obtained from $r$ by removing a finite number of initial unit
delays.

We also define \emph{divergence} coinductively by the rule
\[
\infer={\delay{r} \divg}{
r \divg
}
\]

Now \emph{termination-sensitive weak bisimilarity} is defined
coinductively by the rules
\[
\begin{array}{c}
\infer={r \approx r_*}{
  r \dn r'
  &
  r' \cong r_*'
  &
  r_* \dn r_*'
}
\\[1ex]
\infer={\delay{r} \approx \delay{r_*}}{
  r \approx r_*
}
\end{array}
\]
using an auxiliary predicate defined as a disjunction by the rules
\[
\infer{\ret{\st} \cong \ret{\st}}{
}
\quad
\infer{r_0 + r_1 \cong r_{0*} + r_{1*}}{
  r_0 \approx r_{0*}
  &
  r_1 \approx r_{1*}
}
\quad
\infer{\yield{s}~\st \cong \yield{s}~\st}{
}
\]

While it might seem reasonable to replace the second rule in the
definition of weak bisimilarity by 
\[
\infer={r \approx r_*}{
  r \divg  
  &
  r_* \divg
}
\]
it is actually not a good idea in a constructive setting (classically
one gets an equivalent definition).  Constructively, it is not the
case that any resumption would either converge or diverge (it takes
the lesser principle of omniscience, a weak instance of excluded
middle to prove this). Therefore, we would not be able to prove weak
bisimilarity reflexive.

\section{Functional-style semantics}
\label{sec:funct}

Since we collect the possible executions of a statement into a single
computation tree and divergence is represented by infinite delays,
evaluation of the big-step semantics is deterministic (up to strong
bisimilarity of big-step resumptions) and total: on one hand, for any
$r$, $r_*$, if $\exec{s}{\st}{r}$ and $\exec{s}{\st}{r_*}$, then
$\bism{r}{r_*}$, and on the other, there exists $r$ such that
$\exec{s}{\st}{r}$. This means that the evaluation relation can be
turned into a function.  As a result, from the constructive point of
view, evaluations can not only be checked, but also computed---which
is only good of course.

Here is an equational specification of this function that can be
massaged into an honest definition by structural corecursion.

\noindent Evaluation:
\[
\begin{array}{lllcl}
\eval & (x := e) & \st & = & \delay{(\ret{\st[x \mapsto \sem{e}{\st}]})} \\
\eval &\Skip\,   & \st & = & \ret{\st} \\
\eval &(s_0; s_1) & \st & = &  \evalseq~ s_1~ (\eval~ s_0~ \st) \\
\eval &(\Ifthenelse{e}{s_t}{s_f}) & \st & = & 
              \mathit{if}~\sem{e}{\st}~\mathit{then}~ 
          \delay{(\yield{s_t}~\st)} ~\mathit{else}~ \delay{(\yield{s_t}~ \st)} \\
\eval &(\While{e}{s_t}) & \st & = & 
              \mathit{if}~\sem{e}{\st}~\mathit{then}~ 
          \delay{(\yield{(s_t ; \While{e}{s_t})}~ \st)} ~\mathit{else}~ \delay{(\ret{\st})}
\\
\eval &(s_0 \parallel s_1) & \st & = &  \evalparR~ s_1~ (\eval~ s_0~ \st)
                                     + \evalparL~ s_0~ (\eval~ s_1~ \st) \\
\eval &(\Atomic{s}) & \st & = & \close~ (\eval~ s~ \st) \\
\eval &(\Await{e}{s}) & \st & = & \mathit{if}~\sem{e}{\st}~
              \mathit{then}~\delay{(\close~ (\eval~ s~ \st))}~
              \mathit{else}~\delay{(\yield{(\Await{e}{s})}~ \st)}
 \\
\end{array}
\]

\noindent Sequential extension of evaluation:
\[
\begin{array}{lllcl}
\evalseq & s & (\ret{\st}) & = & \yield{s}~\st \\
\evalseq & s & (\delay{r}) & = & \delay{(\evalseq~ s~ r)} \\
\evalseq & s & (r_0 + r_1) & = & \evalseq~ s~ r_0 + \evalseq~ s~ r_1 \\
\evalseq & s & (\yield{s_0}~\st) & = & \yield{(s_0 ; s)}~\st
\end{array}
\]

\noindent Parallel extension of evaluation:
\[
\begin{array}{lllcl}
\evalparR & s & (\ret{\st}) & = & \yield{s}~\st \\
\evalparR & s & (\delay{r}) & = & \delay{(\evalparR~ s~ r)} \\
\evalparR & s & (r_0 + r_1) & = & \evalparR~ s~ r_0 + \evalparR~ s~ r_1 \\
\evalparR & s & (\yield{s_0}~\st) & = & \yield{(s_0 \parallel s)}~\st \\[1ex]
\evalparL & s & (\ret{\st}) & = & \yield{s}~\st \\
\evalparL & s & (\delay{r}) & = & \delay{(\evalparL~ s~ r)} \\
\evalparL & s & (r_0 + r_1) & = & \evalparL~ s~ r_0 + \evalparL~ s~ r_1 \\
\evalparL & s & (\yield{s_1}~\st) & = & \yield{(s \parallel s_1)}~\st
\end{array}
\]

\noindent Closing a resumption:
\[
\begin{array}{llcl}
\close & (\ret{\st}) & = & \ret{\st} \\
\close & (\delay{r}) & = & \delay{(\close~r)} \\
\close & (r_0 + r_1) & = & \close~ r_0 + \close~ r_1 \\
\close & (\yield{s}~ \st) & = & \delay{(\close~(\eval~ s~ \st))} 
\end{array}
\]

Functional and relational evaluation of the big-step semantics agree:
$\bism{\eval~ s~ \st}{r}$ iff $\exec{s}{\st}{r}$.

Similarly, evaluation of the giant-step semantics is deterministic and
total and can be turned into a function.

\noindent Evaluation:
\[
\begin{array}{lllcl}
\geval & (x := e) & \st & = & \delay{(\ret{\st[x \mapsto \sem{e}{\st}]})} \\
\geval &\Skip\,   & \st & = & \ret{\st} \\
\geval &(s_0; s_1) & \st & = &  \gevalseq~ s_1~ (\geval~ s_0~ \st) \\
\geval &(\Ifthenelse{e}{s_t}{s_f}) & \st & = & 
              \mathit{if~}\sem{e}{\st}\mathit{~then~} 
          \delay{(\yield{(\geval~ s_t)}~\st)} 
         \mathit{~else~} \delay{(\yield{(\geval~ s_t)}~ \st)} \\
\geval &(\While{e}{s_t}) & \st & = & 
              \mathit{if}~\sem{e}{\st}~\mathit{then}~ 
          \delay{(\yield{(\gevalseq~ (\While{e}{s_t}) \circ \geval~ s_t)}~ \st)}
\\
& & & & \hspace*{1.2cm}
 ~\mathit{else}~ \delay{(\ret{\st})}
\\
\geval &(s_0 \parallel s_1) & \st & = &  
                                 \gmergeR~ (\geval~ s_1)~ (\geval~ s_0~ \st)
                               + \gmergeL~ (\geval~ s_0)~ (\geval~ s_1~ \st) \\
\geval &(\Atomic{s}) & \st & = & \gclose~ (\geval~ s~ \st) \\
\geval &(\Await{e}{s}) & \st & = & \mathit{if}~\sem{e}{\st}~
              \mathit{then}~\delay{(\gclose~ (\geval~ s~ \st))}
\\
& & & & \hspace*{1.2cm}
              ~\mathit{else}~\delay{(\yield{(\geval~ (\Await{e}{s}))}~ \st)}
 \\
\end{array}
\]

\noindent Sequential extension of evaluation:
\[
\begin{array}{lllcl}
\gevalseq & s & (\ret{\st}) & = & \yield{(\geval~ s)}~\st \\
\gevalseq & s & (\delay{r}) & = & \delay{(\gevalseq~ s~ r)} \\
\gevalseq & s & (r_0 + r_1) & = & \gevalseq~ s~ r_0 + \gevalseq~ s~ r_1 \\
\gevalseq & s & (\yield{k}~\st) & = & \yield{(\gevalseq~ s \circ k)}~\st
\end{array}
\]

\noindent Merge of a continuation into a resumption:
\[
\begin{array}{lllcl}
\gmergeR & k & (\ret{\st}) & = & \yield{k}~\st \\
\gmergeR & k & (\delay{r}) & = & \delay{(\gmergeR~ k~ r)} \\
\gmergeR & k &  (r_0 + r_1) & = & \gmergeR~ k~ r_0 + \gmergeR~ k~ r_1 \\
\gmergeR & k & (\yield{k_0}~\st) & = & \yield{(\lambda \st'.~ \gmergeR~ k~ (k_0~ \st') + \gmergeL~ k_0~  (k~ \st'))}~\st \\[1ex]
\gmergeL & k & (\ret{\st}) & = & \yield{k}~\st \\
\gmergeL & k & (\delay{r}) & = & \delay{(\gmergeL~ k~ r)} \\
\gmergeL & k &  (r_0 + r_1) & = & \gmergeL~ k~ r_0 + \gmergeL~ k~ r_1 \\
\gmergeL & k & (\yield{k_1}~\st) & = & \yield{(\lambda \st'.~ \gmergeR~ k_1~ (k~ \st') + \gmergeL~ k~  (k_1~ \st'))}~\st \\[1ex]
\end{array}
\]

\noindent Closing a resumption:
\[
\begin{array}{llcl}
\gclose & (\ret{\st}) & = & \ret{\st} \\
\gclose & (\delay{r}) & = & \delay{(\gclose~ r)} \\
\gclose & (r_0 + r_1) & = & \gclose~ r_0 + \gclose~ r_1 \\
\gclose & (\yield{k}~ \st) & = & \delay{(\gclose{(k~ \st)})} 
\end{array}
\]

Functional and relational evaluation of the giant-step semantics
agree: $\gbism{\geval~ s~ \st}{r}$ iff $\gexec{s}{\st}{r}$.

The reduction relation of the small-step semantics is also
deterministic and total and can thus be turned into a function. We
refrain from spelling out the details here.

\section{Trace-based semantics}
\label{sec:traces}

A trace-based big-step semantics is obtained from the resumption-based
big-step semantics straightforwardly by removing the + constructor of
resumptions, splitting the evaluation rule for $\parallel$ into two
rules (thereby turning evaluation nondeterministic) and removing the
rules for + in the definitions of extended evaluations and closing.
Because of the nondeterminism, trace-based evaluation cannot be turned
into a function. But it is still total, as any scheduling leads to a
valid trace. Differently from standard inductive big-step semantics,
divergence from endless work or waiting does not lead to a ``lost
trace''.

In detail, the different ingredients of the semantics are defined as
follows.

\noindent Traces:
\[
\infer={ \ret{\st} : \trace}{ \st : \state}
\quad
\infer={
  \delay{t} : \trace
}{
  t : \trace
}
\quad
\infer={
  \yield{s}~\st : \trace
}{
  s : \stmt
  & 
  \st : \state
}
\]

\noindent Evaluation:
\[
\small
\begin{array}{c}
\infer={\exec{\Assign{x}{e}}{\st}{\delay{(\ret{\update{\st}{x}{\sem{e}{\st}}})}
}
}{}
\qquad
\infer={\exec{\Skip}{\st}{\ret{\st}}
}{}
\quad
\infer={
  \exec{\Seq{s_0}{s_1}}{\st}{t'}
}{
  \exec{s_0}{\st}{t}
  &\execseq{s_1}{t}{t'}
}
\\[1ex]
\infer={
  \exec{\Ifthenelse{e}{s_t}{s_f}}{\st}{\delay{(\yield{s_t}~\st})}
}{
  \istrue{\st}{e}
}
\quad
\infer={
  \exec{\Ifthenelse{e}{s_t}{s_f}}{\st}{\delay{(\yield{s_f}~\st})}
}{
  \isfalse{\st}{e}
}
\\[1ex]
\infer={
  \exec{\While{e}{s_t}}{\st}{\delay{(\yield{(s_t; \While{e}{s_t})}~\st)}}
}{
  \istrue{\st}{e}
}
\quad
\infer={
  \exec{\While{e}{s_t}}{\st}{\delay{(\ret{\st})}}
}{
  \isfalse{\st}{e}
}
\end{array}
\]
\[
\small
\begin{array}{c}
\infer={
  \exec{s_0 \parallel s_1}{\st}{t'}
}{
  \exec{s_0}{\st}{t}
  &
  \execparR{s_1}{t}{t'}
}
\quad
\infer={
  \exec{s_0 \parallel s_1}{\st}{t'}
}{
  \exec{s_1}{\st}{t}
  &
  \execparL{s_0}{t}{t'}
}
\quad
\infer={
  \exec{\Atomic{s}}{\st}{t'}
}{
  \exec{s}{\st}{t}
  &
  \atomize{t}{t'}
}
\\[1ex]
\infer={
  \exec{\Await{e}{s}}{\st}{\delay{t'}}
}{
  \st \models e
  &
  \exec{s}{\st}{t}
  &
  \atomize{t}{t'}
}
\quad
\infer={
  \exec{\Await{e}{s}}{\st}{\delay{(\yield{(\Await{e}{s})}~\st)}}
}{
  \st \not\models e
}
\end{array}
\]

\noindent Sequential extension of evaluation:
\[
\small
\infer={
  \execseq{s}{\ret{\st}}{\yield{s}~\st}
}{
}
\quad
\infer={
  \execseq{s}{\delay{t}}{\delay{t'}}
}{
  \execseq{s}{t}{t'}
}
\quad
\infer={
  \execseq{s}{\yield{s_0}~\st}{\yield{(s_0 ; s)}~\st}
}{
}
\]

\noindent Parallel extension of evaluation:
\[
\small
\begin{array}{c}
\infer={
  \execparR{s}{\ret{\st}}{\yield{s}~\st}
}{
}
\quad
\infer={
  \execparR{s}{\delay{t}}{\delay{t'}}
}{
  \execparR{s}{t}{t'}
}
\quad
\infer={
  \execparR{s}{\yield{s_0}~\st}{\yield{(s_0 \parallel s)}~\st}
}{
}
\\[1ex]
\infer={
  \execparL{s}{\ret{\st}}{\yield{s}~\st}
}{
}
\quad
\infer={
  \execparL{s}{\delay{t}}{\delay{t'}}
}{
  \execparL{s}{t}{t'}
}
\quad
\infer={
  \execparL{s}{\yield{s_1}~\st}{\yield{(s \parallel s_1)}~\st}
}{
}
\end{array}
\]

\noindent Closing a trace:
\[
\small
\infer={
  \atomize{\ret{\st}}{\ret{\st}}
}{
}
\quad
\infer={
  \atomize{\delay{t}}{\delay{t'}}
}{
  \atomize{t}{t'}
}
\quad
\infer={
  \atomize{\yield{s}~\st}{\delay{t'}}
}{
  \exec{s}{\st}{t}
  &
  \atomize{t}{t'}
}
\]

The giant-step case is more interesting. In giant-step
resumptions, we had two kinds of branching: in addition to the binary
branching of $+$, the branching over all states of $\mathit{yield}$.
For a trace-based giant-step semantics, we would like to have a fully
linear concept of traces with neither kind of branching. Evaluation
must then not only ``guess'' which part of a parallel composition gets
to make the first small step, but also which state control is regained
in after suspension.

Accordingly, we define traces without a $+$ constructor. Moreover, we
modify the typing of $\mathit{yield}$.
\[
\infer={ \ret{\st} : \gtrace}{ \st : \state}
\quad
\infer={
  \delay{t} : \gtrace
}{
  t : \gtrace
}
\quad
\infer={
  \yield{(\st', t)}~\st : \gtrace
}{
  \st': \state 
  &
  t : \gtrace
  & 
  \st : \state
}
\]
The idea is to have the trace $\yield{(\st', t)}~\st$ to stand for a
computation that is suspended in state $\st$. Control is returned to
it in state $\st'$ and then it continues as recorded in trace $t$.

Evaluation is defined as for the resumption-based semantics, but there
are two rules for parallel composition and in the rules where a
$\mathit{yield}$ trace is produced $\st'$ is quantified existentially
in premises rather than universally.
\[
\small
\begin{array}{c}
\infer={\gexec{\Assign{x}{e}}{\st}{\delay{(\ret{\update{\st}{x}{\sem{e}{\st}}})}
}
}{}
\\[1ex]
\infer={\gexec{\Skip}{\st}{\ret{\st}}
}{}
\quad
\infer={
  \gexec{\Seq{s_0}{s_1}}{\st}{t'}
}{
  \gexec{s_0}{\st}{t}
  &\gexecseq{s_1}{t}{t'}
}
\\[1ex]
\infer={
  \gexec{\Ifthenelse{e}{s_t}{s_f}}{\st}{\delay{(\yield{(\st', t)}~\st})}
}{
  \istrue{\st}{e}
  &
  \gexec{s_t}{\st'}{t}
}
\quad
\infer={
  \gexec{\Ifthenelse{e}{s_t}{s_f}}{\st}{\delay{(\yield{(\st', t)}~\st})}
}{
  \isfalse{\st}{e}
  &
  \gexec{s_f}{\st'}{t}
}
\\[1ex]
\infer={
  \gexec{\While{e}{s_t}}{\st}{\delay{(\yield{(\st', t)}~\st)}}
}{
  \istrue{\st}{e}
  &
  \gexec{s_t}{\st'}{t}
  &
  \gexecseq{\While{e}{s_t}}{t}{t'} 
}
\quad
\infer={
  \gexec{\While{e}{s_t}}{\st}{\delay{(\ret{\st})}}
}{
  \isfalse{\st}{e}
}
\\[1ex]
\infer={
  \gexec{s_0 \parallel s_1}{\st}{t'}
}{
  \gexec{s_0}{\st}{t_0}
  &
  \gexec{s_1}{\st'}{t_1}
  &
  \gexecparR{(\st', t_1)}{t_0}{t'}
}
\quad
\infer={
  \gexec{s_0 \parallel s_1}{\st}{t'}
}{
  \gexec{s_1}{\st}{t_1}
  &
  \gexec{s_0}{\st'}{t_0} 
  &
  \gexecparL{(\st', t_0)}{t_1}{t'}
}
\\[1ex]
\infer={
  \gexec{\Atomic{s}}{\st}{t'}
}{
  \gexec{s}{\st}{t}
  &
  \gatomize{t}{t'}
}
\\[1ex]
\infer={
  \gexec{\Await{e}{s}}{\st}{\delay{t'}}
}{
  \istrue{\st}{e}
  &
  \gexec{s}{\st}{t}
  &
  \gatomize{t}{t'}
}
\quad
\infer={
  \gexec{\Await{e}{s}}{\st}{\delay{(\yield{(\st', t)}~\st)}}
}{
  \isfalse{\st}{e}
  &
  \gexec{\Await{e}{s}}{\st'}{t}
}
\\[1ex]
\end{array}
\]

Similar considerations apply to sequential extension of evaluation---
\[
\small
\infer={
  \gexecseq{s}{\ret{\st}}{\yield{(\st', t)}~\st}
}{
  \gexec{s}{\st'}{t}
}
\quad
\infer={
  \gexecseq{s}{\delay{t}}{\delay{t'}}
}{
  \gexecseq{s}{t}{t'}
}
\quad
\infer={
  \gexecseq{s}{\yield{(\st', t)}~\st}{\yield{(\st', t')}~\st}
}{
  \gexecseq{s}{t}{t'}  
}
\]
---and to merging a continuation into a trace---
\[
\small
\begin{array}{c}
\infer={
  \gexecparR{k}{\ret{\st}}{\yield{k}~\st}
}{
}
\quad
\infer={
  \gexecparR{k}{\delay{t}}{\delay{t'}}
}{
  \gexecparR{k}{t}{t'}
}
\quad
\infer={
  \gexecparR{k}{\yield{(\st', t)}~\st}{\yield{(\st', t')}~\st}
}{
  \gexecparR{k}{t}{t'}
}
\quad
\infer={
  \gexecparR{(\st', t)}{\yield{k}~\st}{\yield{(\st', t')}~\st}
}{
  \gexecparL{k}{t}{t'}
}\\[1ex]
\infer={
  \gexecparL{k}{\ret{\st}}{\yield{k}~\st}
}{
}
\quad
\infer={
  \gexecparL{k}{\delay{t}}{\delay{t'}}
}{
  \gexecparL{k}{t}{t'}
}
\quad
\infer={
  \gexecparL{(\st', t)}{\yield{k}~\st}{\yield{(\st', t')}~\st}
}{
  \gexecparR{k}{t}{t'}
}
\quad
\infer={
  \gexecparL{k}{\yield{(\st', t)}~\st}{\yield{(\st', t')}~\st}
}{
  \gexecparL{k}{t}{t'}
}
\end{array}
\]
Closing a trace is defined as follows. Closing a $\mathit{yield}$
trace can only succeed if the control release and grab states
coincide, i.e., the grab state has been guessed correctly for the
closed-system situation.
\[
\small
\infer={
  \gatomize{\ret{\st}}{\ret{\st}}
}{
}
\quad
\infer={
  \gatomize{\delay{t}}{\delay{t'}}
}{
  \gatomize{t}{t'}
}
\quad
\infer={
  \gatomize{\yield{(\st, t)}~\st}{\delay{t'}}
}{
  \gatomize{t}{t'}
}
\]

\section{Conclusion}
\label{sec:concl}

We have shown that, with coinductive denotations and coinductive
evaluation, it is possible to give simple and meaningful big-step
descriptions of semantics of languages with concurrency.  The key
ideas remain the same as in the purely sequential case. Most
importantly, due care must be taken of the possibilities of
divergence. In particular, even diverging loops or await statements
must be productive (by growing resumptions or traces by unit delays).
Finite delays can then be equated by a suitable notion of weak
bisimilarity.

Although we could not delve into this topic here, our definitions and
proofs benefit heavily from the fact that the datatype of resumptions
is a monad, in fact a completely iterative monad, and moreover a free
one (as long as we equate only strongly bisimilar resumptions).

With Wolfgang Ahrendt and Keiko Nakata, we have devised a
coinductive big-step semantics for ABS, an exploratory
object-oriented language with an intricate concurrency model, 
developed in the FP7 ICT project HATS. ABS has cooperative scheduling
of tasks (method invocations) communicating via shared memory (fields)
within every object and preemptive scheduling of objects communicating
via asynchronous method calls and futures. This work will be reported
elsewhere.

\paragraph{Acknowledgements}

This work was first presented at the SEFM 2011 Summer School in
Montevideo in November 2011. I thank the organizers for the
invitation.

This research was supported by the EU FP7 ICT project HATS, the ERDF
financed CoE project EXCS, the Estonian Science Foundation grant
no.~9475 and the Estonian Ministry of Education and Research
target-financed research theme no.~0140007s12.

\newcommand{\doi}[1]{\href{http://dx.doi.org/#1}{doi: #1}}

\appendix

\section{Resumption-based semantics for cooperative scheduling}
\label{sec:coop}

Here we give the syntax of Section~\ref{sec:syntax} a cooperative
scheduling interpretation.

It might be argued that this interpretation is more foundational than
the pre-emptive scheduling interpretation---all control release is
explicit and is only due to $\mathsf{await}$ statements. Hence all
$\mathit{yield}$s stem from evaluation of $\mathsf{await}$ statements.

\subsection{Big-step semantics}

\noindent Evaluation:
\[
\small
\begin{array}{c}
\infer={\exec{\Assign{x}{e}}{\st}{\delay{(\ret{\update{\st}{x}{\sem{e}{\st}}})}
}
}{}
\qquad
\infer={\exec{\Skip}{\st}{\ret{\st}}
}{}
\quad
\infer={
  \exec{\Seq{s_0}{s_1}}{\st}{r'}
}{
  \exec{s_0}{\st}{r}
  &\execseq{s_1}{r}{r'}
}
\\[1ex]
\infer={
  \exec{\Ifthenelse{e}{s_t}{s_f}}{\st}{\delay{r}}
}{
  \istrue{\st}{e}
  &
  \exec{s_t}{\st}{r}
}
\quad
\infer={
  \exec{\Ifthenelse{e}{s_t}{s_f}}{\st}{\delay{r}}
}{
  \isfalse{\st}{e}
  &
  \exec{s_f}{\st}{r}
}
\\[1ex]
\infer={
  \exec{\While{e}{s_t}}{\st}{\delay{r'}}
}{
  \istrue{\st}{e}
  &
  \exec{s_t}{\st}{r}
  &
  \execseq{\While{e}{s_t}}{r}{r'}
}
\quad
\infer={
  \exec{\While{e}{s_t}}{\st}{\delay{(\ret{\st})}}
}{
  \isfalse{\st}{e}
}
\end{array}
\]
\[
\small
\begin{array}{c}
\infer={
  \exec{s_0 \parallel s_1}{\st}{r_0' + r_1'}
}{
  \exec{s_0}{\st}{r_0}
  &
  \execparR{s_1}{r_0}{r_0'}
  &
  \exec{s_1}{\st}{r_1}
  &
  \execparL{s_0}{r_1}{r_1'}
}
\quad
\infer={
  \exec{\Atomic{s}}{\st}{r'}
}{
  \exec{s}{\st}{r}
  &
  \atomize{r}{r'}
}
\\[1ex]
\infer={
  \exec{\Await{e}{s}}{\st}{\delay{r'}}
}{
  \st \models e
  &
  \exec{s}{\st}{r}
  &
  \atomize{r}{r'}
}
\quad
\infer={
  \exec{\Await{e}{s}}{\st}{\delay{(\yield{(\Await{e}{s})}~\st)}}
}{
  \st \not\models e
}
\end{array}
\]

\noindent Sequential extension of evaluation:
\[
\small
\infer={
  \execseq{s}{\ret{\st}}{r}
}{
  \exec{s}{\st}{r}
}
\quad
\infer={
  \execseq{s}{\delay{r}}{\delay{r'}}
}{
  \execseq{s}{r}{r'}
}
\quad
\infer={
  \execseq{s}{r_0 + r_1}{r_0' + r_1'}
}{
  \execseq{s}{r_0}{r_0'}
  &
  \execseq{s}{r_1}{r_1'}
}
\quad
\infer={
  \execseq{s}{\yield{s_0}~\st}{\yield{(s_0 ; s)}~\st}
}{
}
\]

\noindent Parallel extension of evaluation:
\[
\small
\begin{array}{c}
\infer={
  \execparR{s}{\ret{\st}}{r}
}{
  \exec{s}{\st}{r}
}
\quad
\infer={
  \execparR{s}{\delay{r}}{\delay{r'}}
}{
  \execparR{s}{r}{r'}
}
\quad
\infer={
  \execparR{s}{r_0 + r_1}{r_0' + r_1'}
}{
  \execparR{s}{r_0}{r_0'}
  &
  \execparR{s}{r_1}{r_1'}
}
\quad
\infer={
  \execparR{s}{\yield{s_0}~\st}{\yield{(s_0 \parallel s)}~\st}
}{
}
\\[1ex]
\infer={
  \execparL{s}{\ret{\st}}{r}
}{
  \exec{s}{\st}{r}
}
\quad
\infer={
  \execparL{s}{\delay{r}}{\delay{r'}}
}{
  \execparL{s}{r}{r'}
}
\quad
\infer={
  \execparL{s}{r_0 + r_1}{r_0' + r_1'}
}{
  \execparL{s}{r_0}{r_0'}
  &
  \execparL{s}{r_1}{r_1'}
}
\quad
\infer={
  \execparL{s}{\yield{s_1}~\st}{\yield{(s \parallel s_1)}~\st}
}{
}
\end{array}
\]

\noindent Closing a resumption:
\[
\small
\infer={
  \atomize{\ret{\st}}{\ret{\st}}
}{
}
\quad
\infer={
  \atomize{\delay{r}}{\delay{r'}}
}{
  \atomize{r}{r'}
}
\quad
\infer={
  \atomize{r_0 + r_1}{r_0' + r_1'}
}{
  \atomize{r_0}{r_0'}
  &
  \atomize{r_1}{r_1'}
}
\quad
\infer={
  \atomize{\yield{s}~\st}{\delay{r'}}
}{
  \exec{s}{\st}{r}
  &
  \atomize{r}{r'}
}
\]

\subsection{Giant-step semantics}

\noindent Evaluation:
\[
\small
\begin{array}{c}
\infer={\gexec{\Assign{x}{e}}{\st}{\delay{(\ret{\update{\st}{x}{\sem{e}{\st}}})}
}
}{}
\quad
\infer={\gexec{\Skip}{\st}{\ret{\st}}
}{}
\quad
\infer={
  \gexec{\Seq{s_0}{s_1}}{\st}{r'}
}{
  \gexec{s_0}{\st}{r}
  &\gexecseq{s_1}{r}{r'}
}
\\[1ex]
\infer={
  \gexec{\Ifthenelse{e}{s_t}{s_f}}{\st}{\delay{r}}
}{
  \istrue{\st}{e}
  &
  \gexec{s_t}{\st}{r}
}
\quad
\infer={
  \gexec{\Ifthenelse{e}{s_t}{s_f}}{\st}{\delay{r}}
}{
  \isfalse{\st}{e}
  &
  \gexec{s_f}{\st}{r}
}
\\[1ex]
\infer={
  \gexec{\While{e}{s_t}}{\st}{\delay{r'}}
}{
  \istrue{\st}{e}
  &
  \gexec{s_t}{\st}{r}
  &
  \gexecseq{\While{e}{s_t}}{r}{r'} 
}
\quad
\infer={
  \gexec{\While{e}{s_t}}{\st}{\delay{(\ret{\st})}}
}{
  \isfalse{\st}{e}
}
\\[1ex]
\infer={
  \gexec{s_0 \parallel s_1}{\st}{r_0' + r_1'}
}{
  \gexec{s_0}{\st}{r_0}
  &
  \forall \st'.\,  \gexec{s_1}{\st'}{k_1~\st'}
  &
  \gexecparR{k_1}{r_0}{r_0'}
  &
  \gexec{s_1}{\st}{r_1}
  &
  \forall \st'.\, \gexec{s_0}{\st'}{k_0~\st'} 
  &
  \gexecparL{k_0}{r_1}{r_1'}
}
\\[1ex]
\infer={
  \gexec{\Atomic{s}}{\st}{r'}
}{
  \gexec{s}{\st}{r}
  &
  \gatomize{r}{r'}
}
\\[1ex]
\infer={
  \gexec{\Await{e}{s}}{\st}{\delay{r'}}
}{
  \istrue{\st}{e}
  &
  \gexec{s}{\st}{r}
  &
  \gatomize{r}{r'}
}
\quad
\infer={
  \gexec{\Await{e}{s}}{\st}{\delay{(\yield{k}~\st)}}
}{
  \isfalse{\st}{e}
  &
  \forall \st'.\,\, \gexec{\Await{e}{s}}{\st'}{k~ \st'}
}
\\[1ex]
\end{array}
\]

\noindent Sequential extension of evaluation:
\[
\small
\infer={
  \gexecseq{s}{\ret{\st}}{r}
}{
  \gexec{s}{\st}{r}
}
\quad
\infer={
  \gexecseq{s}{\delay{r}}{\delay{r'}}
}{
  \gexecseq{s}{r}{r'}
}
\quad
\infer={
  \gexecseq{s}{r_0 + r_1}{r_0' + r_1'}
}{
  \gexecseq{s}{r_0}{r_0'}
  &
  \gexecseq{s}{r_1}{r_1'}
}
\quad
\infer={
  \gexecseq{s}{\yield{k}~\st}{\yield{k'}~\st}
}{
  \forall \st'.\,\, \gexecseq{s}{k~ \st'}{k'\, \st'}  
}
\]

\noindent Merging a continuation into a resumption:
\[
\small
\begin{array}{c}
\infer={
  \gexecparR{k}{\ret{\st}}{k~\st}
}{
}
\quad
\infer={
  \gexecparR{k}{\delay{r}}{\delay{r'}}
}{
  \gexecparR{k}{r}{r'}
}
\quad
\infer={
  \gexecparR{k}{r_0 + r_1}{r_0' + r_1'}
}{
  \gexecparR{k}{r_0}{r_0'}
  &
  \gexecparR{k}{r_1}{r_1'}
}
\quad
\infer={
  \gexecparR{k}{\yield{k_0}~\st}{\yield{(\lambda \st'.\, k_0'~\st' + k_1'~\st')}~\st}
}{
  \forall \st'.\,\, \gexecparR{k}{k_0~\st'}{k_0'~\st'}
  &
  \forall \st'.\,\, \gexecparL{k_0}{k~\st'}{k_1'~\st'}
}\\[1ex]
\infer={
  \gexecparL{k}{\ret{\st}}{k~\st}
}{
}
\quad
\infer={
  \gexecparL{k}{\delay{r}}{\delay{r'}}
}{
  \gexecparL{k}{r}{r'}
}
\quad
\infer={
  \gexecparL{k}{r_0 + r_1}{r_0' + r_1'}
}{
  \gexecparL{k}{r_0}{r_0'}
  &
  \gexecparL{k}{r_1}{r_1'}
}
\quad
\infer={
  \gexecparL{k}{\yield{k_1}~\st}{\yield{(\lambda \st'.\, k_0'~\st' + k_1'~\st')}~\st}
}{
  \forall \st'.\,\, \gexecparR{k_1}{k~\st'}{k_0'~\st'}
  &
  \forall \st'.\,\, \gexecparL{k}{k_1~\st'}{k_1'~\st'}
}
\end{array}
\]

\noindent Closing a resumption:
\[
\small
\infer={
  \gatomize{\ret{\st}}{\ret{\st}}
}{
}
\quad
\infer={
  \gatomize{\delay{r}}{\delay{r'}}
}{
  \gatomize{r}{r'}
}
\quad
\infer={
  \gatomize{r_0 + r_1}{r_0' + r_1'}
}{
  \gatomize{r_0}{r_0'}
  &
  \gatomize{r_1}{r_1'}
}
\quad
\infer={
  \gatomize{\yield{k}~\st}{\delay{r}}
}{
  \gatomize{k\, \st}{r}
}
\]

\subsection{Small-step semantics}

Reduction:
\[
\small
\begin{array}{c}
\infer{\red{x:=e}{\st}{\delay{(\Skip, \st[x \mapsto \sem{e} \st])}}
}{
}
\\[1ex]
\infer{\red{\Skip}{\st}{\ret{\st}}
}{
}
\\[1ex]
\infer{\red{s_0; s_1}{\st}{c}
}{
  \red{s_0}{\st}{\ret{\st'}}
  &
  \red{s_1}{\st'}{c}
}
\quad
\infer{\red{s_0; s_1}{\st}{\delay{(s_0' ; s_1, \st')}}
}{
  \red{s_0}{\st}{\delay{(s_0',\st')}}
}
\\[1ex]
\infer{\red{s_0; s_1}{\st}{(s_{00} ; s_1,\st_0) + (s_{01}; s_1,\st_1)}
}{
  \red{s_0}{\st}{(s_{00},\st_0) + (s_{01},\st_1)}
}
\quad
\infer{\red{s_0; s_1}{\st}{\yield{(s_0' ; s_1)}~ \st'}
}{
  \red{s_0}{\st}{\yield{s_0'}~\st'}
}
\\[1ex]
\infer{\red{\Ifthenelse{e}{s_t}{s_f}}{\st}{\delay{(s_t, \st)}}
}{
  \st \models e
}
\quad
\infer{\red{\Ifthenelse{e}{s_t}{s_f}}{\st}{\delay{(s_f, \st)}}
}{
 \st \not\models e
}
\\[1ex]
\infer{\red{\While{e}{s_t}}{\st}{\delay{(s_t ; \While{e}{s_t}, \st)}}
}{
 \st \models e
}
\quad
\infer{\red{\While{e}{s_t}}{\st}{\delay{(\Skip, \st)}}
}{
 \st \not\models e
}
\\[1ex]
\infer{\red{s_0 \parallel s_1}{\st}{(s_0 \lpar s_1, \st) + (s_0 \rpar s_1, \st)}
}{
}
\\[1ex]
\infer{\red{s_0 \lpar s_1}{\st}{c}
}{
  \red{s_0}{\st}{\ret{\st'}}
  &
  \red{s_1}{\st'}{c}
}
\quad
\infer{\red{s_0 \lpar s_1}{\st}{\delay{(s_0' \lpar s_1, \st')}}
}{
  \red{s_0}{\st}{\delay{(s_0',\st')}}
}
\\[1ex]
\infer{\red{s_0 \lpar s_1}{\st}{(s_{00} \lpar s_1,\st_0) + (s_{01} \lpar s_1,\st_1)}
}{
  \red{s_0}{\st}{(s_{00},\st_0) + (s_{01},\st_1)}
}
\quad
\infer{\red{s_0 \lpar s_1}{\st}{\yield{(s_0' \parallel s_1)}~ \st'}
}{
  \red{s_0}{\st}{\yield{s_0'}~\st'}
}
\\[1ex]
\infer{\red{s_0 \rpar s_1}{\st}{c}
}{
  \red{s_1}{\st}{\ret{\st'}}
  &
  \red{s_0}{\st'}{c}
}
\quad
\infer{\red{s_0 \rpar s_1}{\st}{\delay{(s_0 \rpar s_1', \st')}}
}{
  \red{s_1}{\st}{\delay{(s_1',\st')}}
}
\\[1ex]
\infer{\red{s_0 \rpar s_1}{\st}{(s_0 \rpar s_1,\st_{10}) + (s_0 \rpar s_1,\st_{11})}
}{
  \red{s_1}{\st}{(s_{10},\st_0) + (s_{11},\st_1)}
}
\quad
\infer{\red{s_0 \rpar s_1}{\st}{\yield{(s_0 \parallel s_1')}~ \st'}
}{
  \red{s_1}{\st}{\yield{s_1'}~\st'}
}
\\[1ex]
\infer{\red{\Atomic{s}}{\st}{\ret{\st'}}
}{
  \red{s_}{\st}{\ret{\st'}}
}
\quad
\infer{\red{\Atomic{s}}{\st}{\delay{(\Atomic{s'}, \st')}}
}{
  \red{s}{\st}{\delay{(s',\st')}}
}
\\[1ex]
\infer{\red{\Atomic{s}}{\st}{(\Atomic{s_0},\st_0) + (\Atomic{s_1},\st_1)}
}{
   \red{s}{\st}{(s_0,\st_0) + (s_1,\st_1)}
}
\quad
\infer{\red{\Atomic{s}}{\st}{\delay{(\Atomic{s'}, \st')}}
}{
  \red{s}{\st}{\yield{s'}~\st'}
}
\\[1ex]
\infer{\red{\Await{e}{s}}{\st}{\delay{(\Atomic{s}, \st)}}
}{
 \st \models e
}
\quad
\infer{\red{\Await{e}{s}}{\st}{\delay{(\Suspend; \Await{e}{s}, \st)}}
}{
 \st \not\models e
}
\\[1ex]
\infer{\red{\Suspend}{\st}{\yield{\Skip}~\st}
}{
}
\end{array}
\]
Here $\Suspend$ is an auxiliary statement form that we need for giving
the reduction rule for $\mathsf{await}$. It releases control
immediately (differently from $\Await{\true}{\Skip}$ which makes
a small internal step first).

\end{document}